\documentclass[letterpaper, 10 pt, conference]{ieeeconf}  

\IEEEoverridecommandlockouts                              
\usepackage{graphicx}
\usepackage{epic}
\usepackage{wrapfig, epsfig}
\usepackage{color}
\usepackage[noadjust]{cite}      
\usepackage{graphicx}  
\usepackage{amsmath,amssymb,epsfig,stmaryrd,amsfonts,rotating}

\IEEEtriggeratref{100}

\newcommand{\dr}{d_r}
\newcommand{\dl}{d_l}
\newcommand{\dlh}{\hat{d}_l}
\newcommand{\BPsmall}{\ensuremath{\text{\tiny BP}}} 
\newcommand{\MAPsmall}{\ensuremath{\text{\tiny MAP}}} 
\newcommand{\naturals}{\ensuremath{\mathbb{N}}}
\title{\LARGE \bf
The Effect of Spatial Coupling on Compressive Sensing
}

\author{Shrinivas~Kudekar and Henry~D.~Pfister
\thanks{S. Kudekar is with the New Mexico Consortium and CNLS, Los Alamos National Laboratory, New Mexico, USA.
        {\tt\small skudekar@lanl.gov}}%
\thanks{H.D. Pfister is with the Department of Electrical and Computer Engineering, Texas A\&M University,
        College Station, TX, USA.
        {\tt\small hpfister@tamu.edu}}%
}

\begin{document}

\maketitle
\thispagestyle{empty}
\pagestyle{empty}

\begin{abstract}
Recently, it was observed that spatially-coupled LDPC code
ensembles approach the Shannon capacity for a class of binary-input memoryless
symmetric (BMS) channels. The fundamental reason for this was attributed to a {\em
threshold saturation} phenomena derived in \cite{KRU10}.  In particular, it was shown
that the belief propagation (BP) threshold of the spatially coupled codes is
equal to the maximum a posteriori (MAP) decoding threshold of the underlying
constituent codes. 
In this sense, the BP threshold is saturated to its maximum value.
Moreover, it has been empirically observed that the same phenomena also occurs
when transmitting over more general classes of BMS channels.

In this paper, we show that the effect of spatial coupling is not restricted to the realm of channel
coding. The effect of coupling also manifests itself in compressed sensing.
Specifically, we show that spatially-coupled measurement matrices have an
improved sparsity to sampling threshold for reconstruction algorithms
based on verification decoding. For BP-based reconstruction algorithms,
this phenomenon is also tested empirically via simulation.
At the block lengths accessible via simulation, the effect is quite small
and it seems that spatial coupling is not providing the gains one might expect.
Based on the threshold analysis, however, we believe this warrants further study.

\end{abstract}


\section{Introduction} This work investigates the effect of {\em spatial
coupling} in compressed sensing.  Spatially-coupled codes are a class of
protograph-based low-density parity-check (LDPC) codes capable of achieving
near capacity performance, under low-complexity belief propagation (BP)
decoding, when transmitting over binary-input memoryless symmetric (BMS)
channels. The history of these codes can be traced back to the work of
Felstr\"om and Zigangirov \cite{FeZ99}, where they were introduced as
convolutional LDPC code ensembles.  There is a considerable literature on
convolutional-like LDPC ensembles.  Variations in their constructions as well as
some analysis can be found in \cite{EnZ99, ELZ99,LTZ01,TSSFC04
,SLCZ04,LSZC05,LSZC10, MPZC08,LFZC09,SPVC06,SPVC09, PISWC10, LeF10}.

The fundamental reason underlying the remarkable performance
was recently discussed in detail in \cite{KRU10} for the
case where the transmission takes place over the binary
erasure channel.  Before we go further, we briefly
explain the construction of spatially-coupled codes;
for more details, see \cite{KRU10}. 
  
\subsection{Spatially coupled codes -- ($\dl, \dr, L$) ensemble} 
\begin{figure}[htp]
\begin{centering}
\input{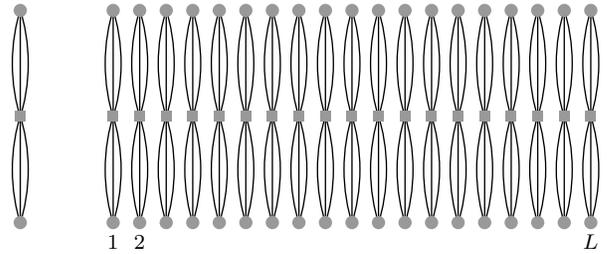}
\caption{On the left is the protograph of a standard $(3, 6)$-regular
ensemble. The graph on the right illustrates a chain of $L$ protographs of the standard $(3, 6)$-regular ensemble
for $L=19$.  These protographs do not interact.}
\label{fig:36protographchain}
\end{centering}
\end{figure}

\begin{figure}[b]
\begin{centering}
\input{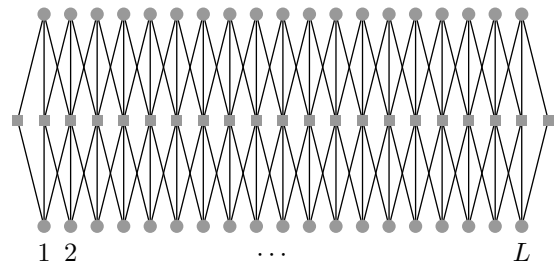}
\caption{A coupled chain of protographs with $L=19$ and $(\dl=3, \dr=6)$. 
 \label{fig:chain}}
\end{centering}
\vspace{-2mm}
\end{figure}

Recall that a regular $(\dl, \dr)$ LDPC
code ensemble can be represented by the protograph (or base graph) as shown in
Fig.~\ref{fig:36protographchain}. Spatially coupled code ensemble, denoted by
$(\dl, \dr, L)$, is constructed by considering a protograph created by taking
multiple copies of the $(\dl,\dr)$ protograph (see the figure on the
right-hand-side in Fig.~\ref{fig:36protographchain}) and connecting them as
shown in the Fig.~\ref{fig:chain}. We stress here that this is only the
protograph. The code is constructed by taking multiple copies of this base graph
and interconnecting them using a random permutation.

For the sake of exposition we consider $(\dl,k\dl)$-regular LDPC ensemble, with $\dl$ odd and $\dlh=(\dl-1)/2\in \naturals$. 
However, coupled codes can
also be constructed starting from any standard ($\dl,\dr$)-regular LDPC code
ensemble \cite{KRU10}.    
To achieve the coupling, connect each protograph to $\dlh$
protographs ``to the left'' and to $\dlh$ protographs ``to the right.''
This is  shown in Figure~\ref{fig:chain} for the case $(\dl=3, \dr=6)$ and $L=9$.
An extra $\dlh$ check nodes are added on each
side to connect the ``overhanging'' edges at the boundary.

There are two main effects resulting from this coupling (see \cite{KRU10} for details): \\
(i) {\em Rate Reduction:} 
Recall that the design rate of the underlying standard $(\dl, \dr=k
\dl)$-regular ensemble is $1-\frac{\dl}{\dr}=\frac{k-1}{k}$. 
The design rate of the corresponding $(\dl, \dr=k \dl, L)$
 ensemble, due to boundary effects, is reduced to
\begin{align}\label{eq:rateloss}
R(\dl, \dr=k\dl, L) 
& = \frac{k-1}{k} - \frac{\dl-1}{kL}.
\end{align} 
(ii) {\em Threshold Increase:}  Let $\epsilon^{\BPsmall}(\dl, \dr, L)$,
$\epsilon^{\MAPsmall}(\dl, \dr, L)$ and $\epsilon^{\MAPsmall}(\dl, \dr)$ denote
the  threshold\footnote{The BP(MAP) threshold, of a fixed ensemble
of codes, denotes the channel parameter value below which the BP(MAP) decoder
succeeds and fails above it.} of the BP decoder for the $(\dl, \dr, L)$ ensemble, MAP
threshold of the $(\dl, \dr, L)$ ensemble and the MAP threshold of the
underlying $(\dl, \dr)$ ensemble, respectively.  Then the main result of
\cite{KRU10} is that, when we transmit spatially coupled codes over the BEC we have, 
$$ \epsilon^{\BPsmall}(\dl, \dr, L) \approx
\epsilon^{\MAPsmall}(\dl, \dr, L) \approx \epsilon^{\MAPsmall}(\dl, \dr).$$ See
\cite{KRU10} for a precise statement of the main theorem. The effect of coupling
can been nicely seen by plotting the EXIT curves for the uncoupled and coupled codes.
This is shown in Fig.~\ref{fig:BECEXIT}. Similar phenomena can also  be empirically observed when
transmitting over more general BMS channels \cite{KMRU10, LMFC10}.   

\subsection{Outline} In this work, we study the effect of spatial coupling
in the problem of compressed sensing. We begin with our compressive 
sensing setup and explain our decoders in the next section. In the same
section, we introduce spatially-coupled measurement matrices. We then develop
the density evolution (DE) equations for the class of decoders which we consider. In
Section~\ref{sec:expt} we perform experiments depicting the effect of spatial
coupling. We conclude with a short discussion of interesting open questions.  

\section{Compressed Sensing}

Compressed sensing (CS) is now one of the most exciting new areas in signal
processing.  It is based on the idea that many real-world signals (e.g., those
sparse in some transform domain) can be reconstructed from a small number of
linear measurements.  This idea originated in the areas of statistics and
signal processing
\cite{Chen-siamscicomp98,Gorodnitsky-sp02,Cotter-sp05,Donoho-it06,Candes-it06},
but is also quite related to previous work in computer science
\cite{Gilbert-soda03,Cormode-ja05,Gilbert-stoc07} and applied mathematics
\cite{Cohen-tr06,Johnson-cm84,Gluskin-mus84}.  CS is also very closely related
to error-correcting codes, and can be seen as source coding using linear codes
over real numbers
\cite{Sarvotham-isit06,Xu-itw07,Zhang-ita08,Sarvotham-tr06,Dai-arxiv08, DMM10,
Donoho-pnas09, Donoho-pnas09supp}. 

The basic problem is to measure and then reconstruct a \emph{signal vector} $x \in \mathbb{R}^N$ using the linear observation model $y=\Phi x + w$, where the matrix $\Phi \in \mathbb{R}^{M \times N}$ is the \emph{measurement matrix} and $w$ is a noise vector.
The signal vector is called \emph{$K$-sparse} if it contains at most $K$ non-zero entries.
If, instead of being identically zero, the other $N-K$ entries are much smaller than the largest $K$ entries, then the vector is called \emph{approximately sparse}.

The main goal of this paper is to study the effect of spatial coupling on the performance of CS systems.
Spatial coupling has been shown to drastically increase the threshold LDPC codes with an asymptotically negligible decrease in the code rate.
The tight connection between low-density parity-check (LDPC) codes and CS means that one would expect spatial coupling to improve message-passing decoders for CS systems.
A more subtle question is whether or not traditional CS decoders based on convex relaxations (e.g., basis pursuit and LASSO) will also benefit from spatial coupling.
In this paper, we compare various standard constructions of measurement matrices with constructions based on spatial coupling.
These tests are performed using a variety of CS decoders and a few signal models.

\begin{figure}[t!]
\begin{centering}
\scalebox{1.6}{\input{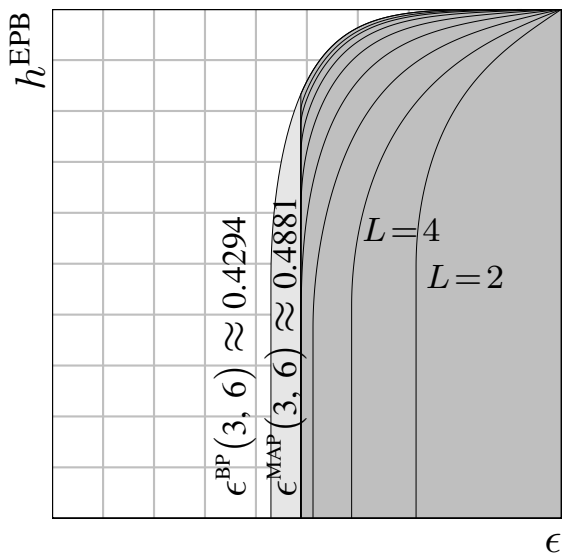}}
\vspace{2mm}
\caption{The figure depicts the BP EXIT curves of the ensemble $(3, 6, L)$ 
for $L=2, 4, 8, 16, 32, 64, 128$, and $256$, when transmitting over the binary erasure channel.
The light/dark gray areas mark the interior of the BP/MAP EXIT function
of the underlying $(3, 6)$-regular ensemble, respectively. For small values of
$L$, the increase in threshold is because of the rate loss. As $L$ increases the
curves keep moving to the left and get ``stuck'' at the MAP threshold of the
underlying regular ensemble.  
\label{fig:BECEXIT}
}
\end{centering}
\end{figure}

\subsection{System Model}

All our CS results rely on sparse measurement matrices.
A variety of such matrices have been considered before in \cite{Sarvotham-isit06,Sarvotham-tr06,Xu-itw07,Zhang-ita08,Zhang-aller08,Berinde-aller08}.
Their main advantage is that they enable a variety of low-complexity reconstruction techniques.
In general, the entries of the measurement matrix are either chosen to be plus/minus one with equal probability, or drawn from a continuous distribution.
The latter provides some benefit when the signal contains only a small set of non-zero values.

Throughout the paper we will consider two kinds of measurement matrices.  The
first type of measurement matrix is generated from the parity-check matrix
sampled uniformly at random from the ensemble of a regular $(\dl, \dr=k\dl)$
LDPC codes.  The sampling ratio, $\delta$, is given by $1/k$.  The second type
of measurement matrix, which we call a {\em spatially coupled measurement
matrix}, will be generated from the parity-check matrix sampled from the $(\dl,
\dr, L)$ ensemble.  In this case, the sampling ratio can be obtained from
\eqref{eq:rateloss} as $\delta=\frac{1}k + \frac{\dl-1}{kL}$.  For finite $L$,
the number of measurements of the spatially coupled measurement matrix is
larger than the corresponding uncoupled matrix. But as $L\to \infty$, the two
measurement matrices have an equal number of measurements.

The signal $x\in \mathbb{R}^N$ is assumed to be drawn i.i.d. from some distribution $f_X (x)$.
The parameter $\epsilon$ determines the fraction non-zero entries in the signal.
To model $\epsilon N$-sparse vectors, one typically assumes that $f_X (x)$ has a delta function at $x=0$ with mass $1-\epsilon$.
For approximately sparse vectors, we use the two-Gaussian mixture model from \cite{Sarvotham-tr06,Baron-sp10}.
In this model, all entries are independent zero-mean Gaussians but a fraction
$\epsilon N$ have variance $\sigma^2_1$ and a fraction $(1-\epsilon)N$ have
variance $\sigma^2_0$ (with $\sigma_1>\sigma_0$).

\subsection{Message Passing Reconstruction}

Message-passing (MP) reconstruction for compressed-sensing systems based on sparse measurement matrices was introduced by Sarvotham, Baron, and Baraniuk in \cite{Sarvotham-isit06,Sarvotham-tr06,Baron-sp10}.
The tight connection between CS and error-correcting codes enabled researchers to quickly analyze other MP reconstruction schemes based on tools from modern coding theory \cite{Xu-itw07,Zhang-ita08,Zhang-aller08,Dai-arxiv08}.
Recently, there has also been some progress in analyzing the performance of
these schemes for approximately sparse signals and noisy observations
\cite{Baron-sp10,Donoho-pnas09,Bayati-arxiv10}.

For sparse measurement matrices, the asymptotic performance of MP
reconstruction can be analyzed (in theory) using density evolution (DE)
\cite{Richardson-it00}.  Indeed, this works well for simplified suboptimal
reconstruction algorithms like the "sudocodes" reconstruction
\cite{Sarvotham-isit06,Zhang-arxiv09}.  For true belief-propagation (BP)
reconstruction, however, numerical evaluation of the DE equations is
intractable.  This means that it is difficult to determine the asymptotic
behavior of a particular sparse ensemble with BP reconstruction.  Recently,
Donoho, Maleki and Montanari \cite{Donoho-pnas09}, \cite{DMM10} proposed an {\em approximate message passing}
(AMP) algorithm for compressed sensing with dense Gaussian measurement
matrices.  For this algorithm, they introduced a variant of DE (known as state
evolution) that provides a precise characterization of its performance.

\subsection{Analytical Setup}\label{sec:verdecoder}

Our analytical results rely on the suboptimal reconstruction technique
introduced for "sudocodes" by Sarvotham, Baron, and Baraniuk
\cite{Sarvotham-isit06}.  The "sudocodes" reconstruction technique falls into
the class of \emph{verification decoders} that was introduced and analyzed by
Luby and Mitzenmacher for LDPC codes over large alphabets \cite{Luby-it05}.  In
this paper, we use the message-passing based implementation of the second (more
powerful) algorithm from their paper and refer to it as LM2
\cite{Zhang-arxiv08}.  The main drawback of this choice is that the analysis
only works for strictly sparse vectors where the measurements are observed
without noise.  The main benefit is that one can analyze its performance
precisely using density evolution (DE) and construct EXIT-like curves to
illustrate the benefits of spatial coupling.  For example, in the large system
limit, this leads to provable sparsity thresholds where reconstruction succeeds
with probability one \cite{Zhang-ita08,Zhang-arxiv09}.

Although the LM2 decoder is unstable in the presence of noise, this does not
mean that its threshold is meaningless in practice.  The LM2 decoder can be
seen as a suboptimal version of list-message passing (LMP) \cite{Zhang-arxiv08}
which itself can be seen as a limiting case of the full belief-propagation (BP)
decoder for CS \cite{Sarvotham-tr06,Baron-sp10}.  Ideally, one would analyze
the BP decoder directly, but performing a DE analysis for decoders which pass
real functions as messages \cite{Sarvotham-tr06,Baron-sp10} remains
intractable.  Still, we expect that a complete analysis of the BP decoder
would show that its expected performance is always better than the LM2 decoder
and that the BP decoder allows stable reconstruction below its sparsity threshold.


\begin{center}\framebox[0.95\columnwidth]{
\begin{minipage}{0.85\columnwidth}
\vspace{1mm}
Verification decoding rules for message-based LM2 decoding in a CS system:
\begin{itemize}
\item (Check Node) The output message on an edge:
\begin{itemize}
\item equals the unique value which satisfies the observation constraint based on all other input edges;
\item this output message is verified if and only if all other input messages are verified.
\end{itemize}

\item (Symbol Node) The output message is:
\begin{itemize}
\item verified and equal to the input message if any input message is verified,
\item verified and equal to 0 if any input message on another edge is equal to 0,
\item verified and equal to the matching value if any two other input messages match,
\item and unverified with a value of 0 otherwise.
\end{itemize}
\end{itemize}
\vspace{1mm}
\end{minipage}}
\end{center}
\vspace{3mm}

Initially, the check nodes with measurement equal to zero, transmit zero on all
their outgoing edges. This arises from the basic property of verification
decoders when we consider the non-zero values to come from a continuous
distribution.  The scheme described above does not guarantee that all verified
symbols are actually correct.  The event that a symbol is verified but
incorrect is called false verification (FV).  If either the non-zero matrix
entries, or the signal values, are drawn from a continuous distribution, then
the FV event has probability zero.

\subsubsection{Protograph Density Evolution for LM2 Decoding}

For generality, we consider a verification-based decoding rules for channels with erasures and errors.
The variables $w,x,y,z$ will be used to denote, respectively, the probability that the message type is erasure (E), incorrect (I), correct and unverified (C), and verified (V).
The DE equations for LM2 decoding of the standard irregular LDPC code ensemble are given in \cite[p. 125]{Luby-it05}.
For the protograph setup (including erasures), we derive the DE equations below.

For a check node of degree $d$, let $w_{i},x_{i},y_{i},z_{i}$ the message-type probability for the $i$th input edge and $w_{i}',x_{i}',y_{i}',z_{i}'$ be the message-type probability for the $i$th output edge.
Then, we have the update rules

\begin{align}\label{eq:checkDE}
w_{i}' & =1-\prod_{j\neq i}(1-w_{j}) \nonumber \\
x_{i}' & =\prod_{j\neq i}(1-w_{j})-\prod_{j\neq i}(1-w_{j}-x_{j}) \nonumber \\
y_{i}' & =\prod_{j\neq i}(y_{j}+z_{j})-\prod_{j\neq i}z_{j} \nonumber \\
z_{i}' & =\prod_{j\neq i}z_{j}
\end{align}
In words, these four disjoint probabilities are for the events: ``at
least one E'', ``at least one E or I minus at least E'', ``all
C or V minus all V'', and ``all V''.

For a bit node of degree $d$, let $w_{i},x_{i},y_{i},z_{i}$ the
message-type probability for the $i$th input edge and $w_{i}',x_{i}',y_{i}',z_{i}'$
be the message-type probability for the $i$th output edge. If $\epsilon$
is the probability of channel error and $p$ is the probability of
channel erasure, then we have the update rules

\begin{align}\label{eq:varDE}
w_{i}' & =p\left(\prod_{j\neq i}\left(w_{j}+x_{j}\right)+\sum_{k\neq
i}y_{k}\prod_{j\neq k,i}\left(w_{j}+x_{j}\right)\right) \nonumber \\
x_{i}' & =\epsilon\left(\prod_{j\neq i}\left(w_{j}+x_{j}\right)+\sum_{k\neq
i}y_{k}\prod_{j\neq k,i}\left(w_{j}+x_{j}\right)\right) \nonumber \\
y_{i}' & =(1-\epsilon-p)\prod_{j\neq i}\left(w_{j}+x_{j}\right) \nonumber \\
z_{i}' & =(1-\epsilon-p)\sum_{k\neq i}y_{k}\prod_{j\neq k,i}\left(w_{j}+x_{j}\right) \nonumber \\
& +\left(1-\prod_{j\neq i}\left(w_{j}+x_{j}\right)-\sum_{k\neq
i}y_{k}\prod_{j\neq k,i}\left(w_{j}+x_{j}\right)\right).
\end{align}
Let $A$ be the event that ``all input edges are E or I except
for at most one C''. In words, these four disjoint probabilities
are for the events: ``channel erased and $A$'', ``channel
error and $A$'', ``channel correct and all input edges are E or
I'', and ``not $A$ or channel correct and $A$''.

We can specialize the above DE equations to the case when we do not have spatial
coupling. More precisely, consider the measurement matrix generated by the regular $(\dl, \dr)$ LDPC ensemble. Then
 for the check node we have,
\begin{align}\label{eq:checkDEreg}
w' & =1-(1-w)^{\dr-1} \nonumber \\
x' & =(1-w)^{\dr-1}-(1-w-x)^{\dr-1} \nonumber \\
y' & =(y+z)^{\dr-1}-z^{\dr-1} \nonumber \\
z' & =z^{\dr-1},
\end{align}
and for the variable node side we have, 
\begin{align}\label{eq:varDEreg}
w' & =p\left((w+x)^{\dl-1}+(\dl-1)y(w+x)^{\dl-2}\right)\nonumber \\
x' & =\epsilon\left((w+x)^{\dl-1}+(\dl-1)y(w+x)^{\dl-2}\right)\nonumber \\
y' & =(1-\epsilon-p)(w+x)^{\dl-1} \nonumber \\
z' & =(1-\epsilon-p)(\dl-1)y(w+x)^{\dl-2} \nonumber \\ &
+\left(1-(w+x)^{\dl-1}-(\dl-1)y(w+x)^{\dl-2}\right).
\end{align}
These equations can be compared to \cite[p. 125]{Luby-it05} by setting $p=0$, $w=0$,
 mapping $x \rightarrow b$ , and mapping $y \rightarrow a$. We remark here that
the channel error probability, $\epsilon$, is equal to the fraction of non-zero
symbols in the signal.


\subsubsection{EXIT-like curves} In theory of iterative codes, the EXIT
function is defined by $H(X_i\vert Y_{\sim i})$ \cite{RiU08}, where $Y_{\sim
i}$ denotes the vector of all observations except $Y_i$. In words, it is the
uncertainty in decoding a bit, when its channel observation is discarded.  In
general one can replace $Y_{\sim i}$ by $\phi_{i,\text{dec}}(Y_{\sim i})$,
where $\phi_{i,\text{dec}}(Y_{\sim i})$ is the extrinsic estimate of any
other decoder (e.g., BP decoder). Such EXIT-like curves are useful visualization
tool in iterative coding theory. They have also been used in iterative coding
theory to provide deep results relating the BP decoder and the optimal MAP
decoder \cite{RiU08}.  It is hence of great interest to visualize our results by plotting EXIT-like
 curves even in the case of compressed sensing. 

According to the rules of LM2 decoder, a variable node is verified, or its
value is perfectly known, when the variable node receives either two or more
$C$ (correct but unverified) messages or one or more $V$ (verified) message. In
this case the EXIT value of the variable node is zero. Thus the EXIT value of a
variable node is proportional to the probability that a variable node is
unverified. In other words, it receives ``all I (incorrect) except at most one C
(correct but unverified)". This is exactly the probability of the event A (cf.
Section~\ref{sec:verdecoder}). To summarize: in our experiments we plot an
EXIT-like curve which is the probability of a variable node being unverified
when we change the sparsity ratio continuously.    

\begin{figure}[t!]
\centering
\hspace{9mm}
\input{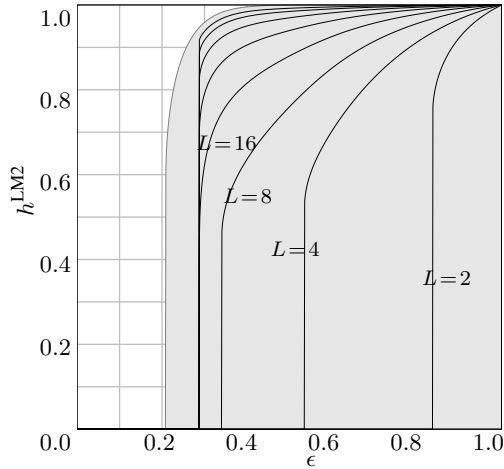}
\caption{\label{fig:exit_4_8}  The figure shows the EXIT-like function for
 the LM2 verification decoder for compressed sensing. The horizontal and vertical axes
correspond to the sparsity ratio, $\epsilon$, and the probability of a variable node being
unverified (at the end of decoding), $h^{\text{LM2}}$, respectively. The light gray curve on the
extreme left corresponds to the case of a measurement matrix generated by a
regular (4,8) LDPC code (i.e., sampling ratio equal to 1/2). The curves in dark correspond to measurement matrices
based on coupled (4,8) LDPC codes of coupling length $L=2,4,8,16,32,64,128,256$.
As $L$ increases the sampling ratio of the coupled measurement matrices go to
$1/2$ (cf. equation \eqref{eq:rateloss}). Similar to channel coding,
we see that the EXIT-like curves keep on moving to the left as $L$ increases.
For large enough $L$, the curves seem to get ``stuck'' at some limiting curve. The phase
transition for the coupled measurement matrix for $L=256$ occurs at $\approx
0.287$, whereas it is $\approx 0.208$ for the uncoupled case.} 
\end{figure}

\section{Experiments}\label{sec:expt}

\subsection{Verification Decoders}

We perform DE analysis of the verification decoder. More precisely, we consider
the DE equations \eqref{eq:checkDE} to \eqref{eq:varDEreg}. In particular we
consider the case when $p=0$, i.e., we have corruption of symbols only via
errors in the channel. 

\subsubsection{EXIT-like curves for fixed sampling ratio} We first consider
measurement matrices generated by the regular $(4,8)$ LDPC code. This fixes the
sampling ratio to $1/2$.  We then fix the sparsity ratio $0<\epsilon<1$ and
run DE equations \eqref{eq:checkDEreg} and \eqref{eq:varDEreg}, till a fixed
point is reached. We then use the fixed point densities to evaluate the
probability that a particular node is unverified. We plot this value on the
vertical axis in Fig.~\ref{fig:exit_4_8}, for different values of $\epsilon$.
In this case, the EXIT-like curve is illustrated by the light gray curve
(leftmost) in Fig.~\ref{fig:exit_4_8}. We observe that below $\epsilon\approx
0.208$, the fixed point is trivial. More precisely, for sparsity ratio
$\epsilon \leq 0.208$, the probability that a node is unverified goes to zero.
This means that, with high probability, the LM2 decoder is able to reconstruct
the signal exactly. For $\epsilon>0.208$, we see that the LM2 decoder fails to
reconstruct the signal and the probability of a variable node being unverified
is non-zero. 

Similar experiment is now done with spatially coupled $(4,8,L)$ measurement
matrices. We run the DE equations given by \eqref{eq:checkDE} and
\eqref{eq:varDE} for different lengths, $L=2,4,8,16,32,64,128,256$. For $L=2$,
the threshold is $\approx 0.837$. The reason for such a large value of the
threshold is the because the sampling ratio is much larger than $0.5$ (cf.
equation \eqref{eq:rateloss}). As $L$ increases the curves move to the left,
which is similar to the effect observed in channel coding (cf.
Fig.~\ref{fig:BECEXIT}). As $L$ increases, the resultant measurement matrix
resembles more and more like the uncoupled $(4,8)$ measurement matrix (locally)
and the sampling ratio also approaches $1/2$.  However, the curves seem to get
stuck at $\epsilon \approx 0.287$. To summarize: for large $L$ we have the sampling
ratio $\delta$ very close to $0.5$ and we observe that the sparsity threshold
is much larger, $\epsilon\approx 0.287$. 

\begin{figure}[t!]
\centering
\scalebox{0.95}{\hspace{9mm} \input{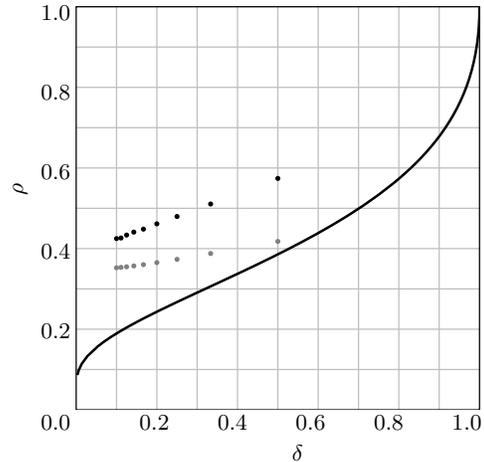}} 
\caption{\label{fig:phase_trans} The figure shows the phase transition between
successful and unsuccessful reconstruction as a function of sampling and sparsity.
The horizontal axis corresponds to the sampling ratio, $\delta =
M/N$. The vertical axis represents the sparsity ratio (normalized by the number of measurements rather than the dimension of the signal), $\rho=\epsilon/\delta=K/M$.
The continuous curve is Donoho's phase transition for weak CS reconstruction using LP \cite{Donoho-dcg06}.
Each light circle (resp. dark circle) corresponds to the phase transition for LM2 reconstruction
when we use an uncoupled (resp. coupled) $(4,4k)$ LDPC code as a measurement matrix with $k=2,3,\ldots,10$.} 
\end{figure}

\subsubsection{Phase transition} We run the DE equations for different values
of the sampling ratio. Let us explain this more precisely. As usual, we 
consider an uncoupled ($\dl, \dr$) measurement matrix generated and its coupled
version, $(\dl, \dr, L)$. We fix $\dl=4$ and consider $\dr= k \dl$ for different values of $k$.
As a consequence, we obtain different values of $\delta=1/k$. We run DE equations
\eqref{eq:checkDE} and \eqref{eq:varDE} for $k=2,3, \ldots, 10$ and for $L=1000$
fixed for all the experiments.

Figure~\ref{fig:phase_trans} shows the results. The light circles correspond to the regular case and the dark circles correspond to the coupled case.
We see that for the sampling ratios considered, the coupled measurement matrices have an improved sparsity-sampling trade-off. 
At the first, the trend of the circles may seem a bit strange.
But, the normalized coordinates of the plot imply that a family of systems achieving a fixed oversampling ratio family would give a horizontal line.

The continuous curve shown is Donoho's phase transition for weak CS reconstruction using LP \cite{Donoho-dcg06}.
This was later identified in \cite{Donoho-pnas09} as the LP threshold for AMP reconstruction of a sparse signal with a worst-case amplitude distribution.
For this reason, the comparison is not really fair.
The LM2 thresholds are for the noiseless measurement of a special class of signals, while LP reconstruction thresholds are for the stable reconstruction of a random signal drawn from any i.i.d. distribution.
Still, this result does highlight the fact that better performance is possible (and might even be achieved by LP reconstruction) in some special cases.

\subsection{BP and LP Reconstruction}

Simulations were also performed for belief-propagation (BP) and linear-programming (LP) reconstruction of coupled and uncoupled measurement matrices.
The initial goal was to choose system parameters that would allow direct comparison with \cite[Fig. 3]{Baron-sp10}.
Unfortunately, the rate loss associated with coupled codes made it difficult to achieve fair comparisons in this case.
Therefore, the tested system uses noiseless measurements to reconstruct signals from the two-Gaussian model with $N=4032$, $\epsilon=0.1$, $\sigma_0 = 0.5$, and $\sigma_1 = 10$.
For the coupled and uncoupled systems, measurement matrices are based on $(\dl,\dr)$ regular codes with $\dl=5,7,9$ and $\dr=2\dl,3\dl,4\dl$.
Non-zero entries in these matrices are randomly chosen to be either $+1$ or $-1$.
These parameters give asymptotic sampling ratios of $\delta=0.25,0.5,0.75$, but the coupled systems have slightly more measurements due to finite-$L$ rate loss.

\begin{figure}
\centering
\includegraphics[width=0.85\columnwidth]{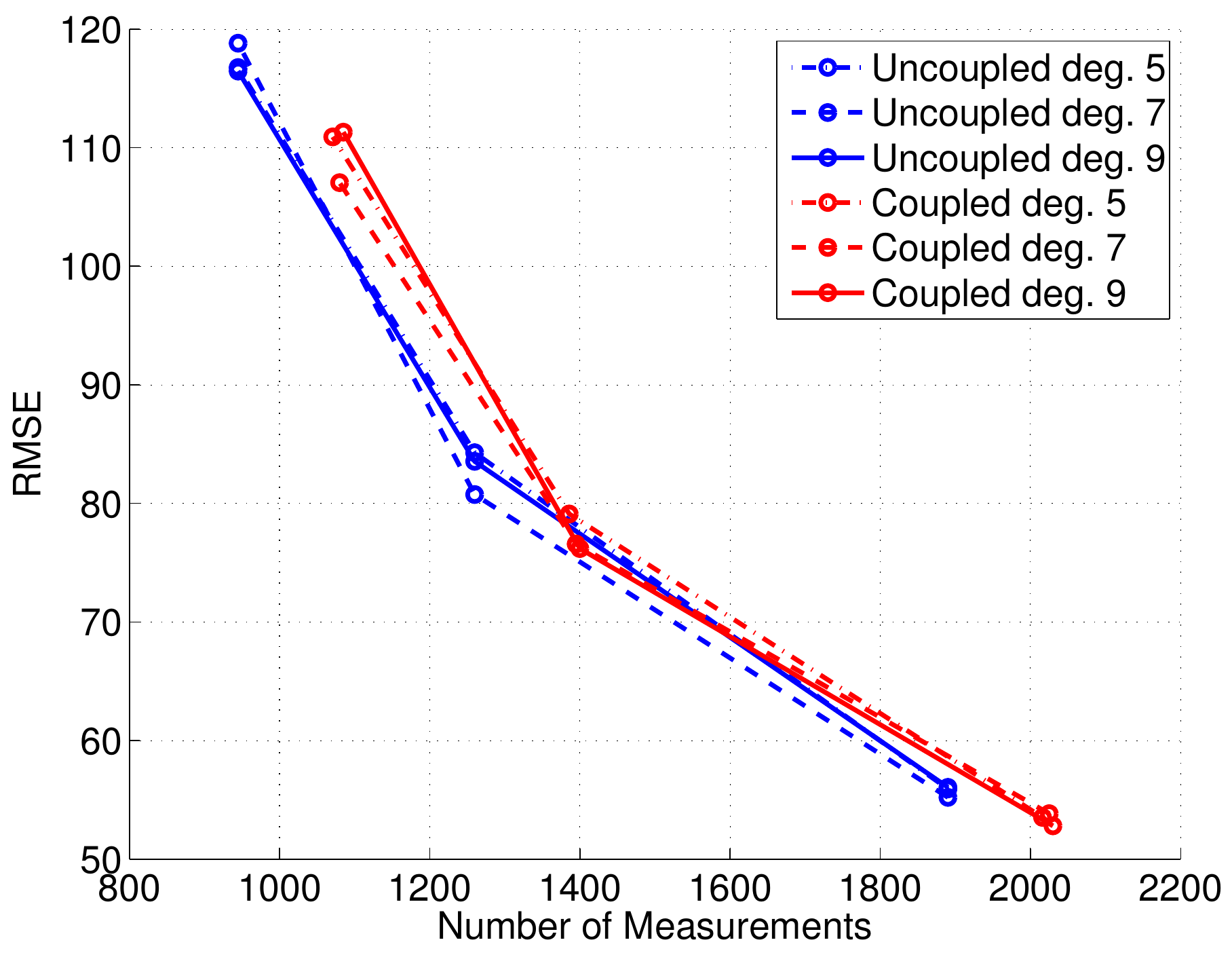}
\caption{ \label{fig:csbp_deg} RMSE reconstruction error for 10 trials of BP reconstruction (with 30 iterations) of signals from the two-Gaussian model with $N=4032$, $\epsilon=0.1$, $\sigma_0 = 0.5$, and $\sigma_1 = 10$.
The measurement matrices used are based on coupled and uncoupled $(\dl,\dr)$-regular codes with $\dl=5,7,9$ and $\dr=2\dl,3\dl,4\dl$.
To fix the block length, the number of coupling stages is chosen to be $L=48 / k $.
These parameters give sampling ratios close to $\delta=0.25,0.5,0.75$, but the coupled systems have slightly more measurements due to the finite-$L$ rate loss.}
\end{figure}

For each setup, the experiment tested a single (randomly constructed)
measurement matrix on the same set of 10 randomly generated signal vectors with
$N=4032$ and $K=403$ (the number of non-zero symbols in the signal).  During
each trial, the BP decoder is run for 30 iterations and the root-mean-square
error (RMSE) is calculated.  The curves in Figure~\ref{fig:csbp_deg} show the
median RMSE (over 10 trials) for the particular parameters.  



\begin{figure}
\centering
\includegraphics[width=0.85\columnwidth]{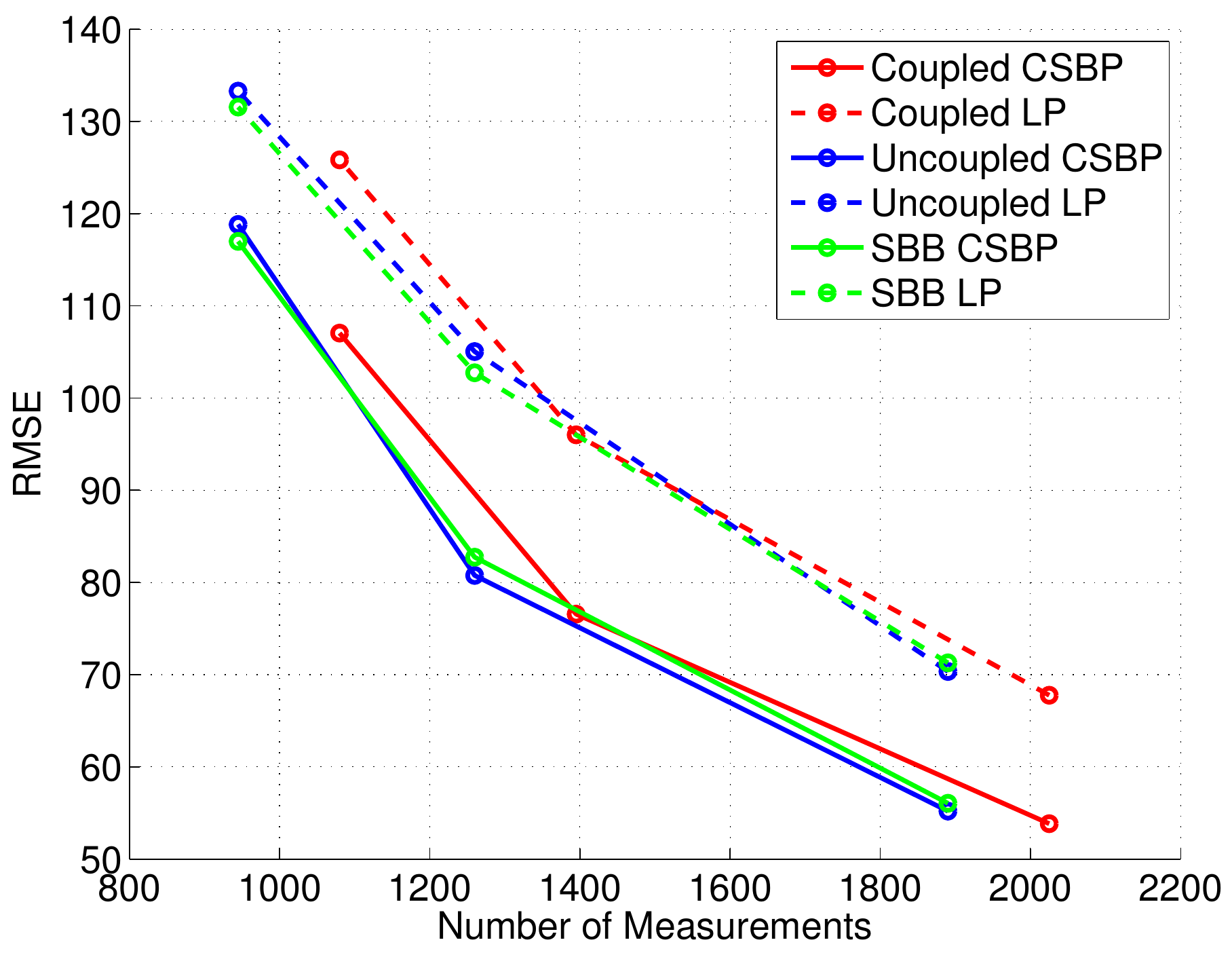}
\caption{\label{fig:csbp_deg7} Median reconstruction error for 10 reconstruction trials of signals from the two-Gaussian model with $N=4032$, $\epsilon=0.1$, $\sigma_0 = 0.5$, and $\sigma_1 = 10$.
Reconstruction is performed either using belief propagation (BP) or linear programming (LP).
The measurement matrices used are based on coupled and uncoupled $(7,7k)$-regular codes with $k=2,3,4$.
To fix the block length, the number of coupling stages is chosen to be $L=48/k$.
These parameters give sampling ratios close to $\delta=0.25,0.5,0.75$, but the coupled systems have slightly more measurements due to the finite-$L$ rate loss.
}
\end{figure}

In Fig.~\ref{fig:csbp_deg} we compare the coupled and uncoupled measurement
matrix for increasing average degrees. In Fig.~\ref{fig:csbp_deg7} we compare
the performance of CSBP and LP decoders when we consider coupled and uncoupled
measurement matrix with variable node degree fixed to 7. The figure also shows
the performance curve when we use the Sarvotham, Barron, Baraniuk (SBB) (see
\cite{Baron-sp10})  ensemble for measuring. 

We remark here that the simulation setup for this section uses $L$ between 24
and 48. The number of coupling stages must be kept somewhat low for two reasons:
(i) to prevent short cycles in the decoding graph and (ii) to reduce the
decoding time as the number of required decoding iterations increases with $L$.
These small values of $L$ result in a slightly larger sampling ratio (i.e., more
measurements) for the coupled measurement matrices as opposed to uncoupled
measurement matrices. This effect is handled correctly on the simulation curves,
but makes direct comparison somewhat difficult.
As a consequence, we observe that there is no appreciable gain by using couplng
in the case when we are using the CSBP decoder.  The performance changes are
very small and do not support a conclusion that spatial-coupling provides large
gains for CS with BP reconstruction.


The basis-pursuit LP reconstruction technique was also tested with each
measurement matrix and signal vector.  The results for symbol degree $\dl=7$ are
shown in Figure~\ref{fig:csbp_deg7}.  From this, one observes that LP
reconstruction benefits even less from spatial coupling.

However, results from the channel
coding setup imply that the performance for moderate $L$ is very close to the
performance when $L$ is very large.  For example, Figures~\ref{fig:BECEXIT} and
\ref{fig:exit_4_8} show that the threshold for $L=16$ has already saturated to
the $L\rightarrow \infty$ threshold.  Nevertheless, experiments
with larger values of $L$ merit further investigation.  


\section{Discussion}

Recently, it was shown in \cite{HMU10} that the effect of coupling is also
observed in many other problems, like the $k$-satisfiability, graph coloring and
Curie-Weiss model of statistical physics. 
In this paper, we found that spatially-coupled measurement matrices have
an improved sparsity-sampling ratio for model used by verification decoding.
This provides evidence that the phenomena of spatial coupling is quite general.
On the other hand, we also observed that spatially-coupled measurement matrices
provide little (if any) gains for the compressed-sensing problem with moderate
blocklengths and belief-propagation reconstruction.

We conclude with possible future research directions.
\begin{enumerate}
\item  It would be interesting to see the effect of spatial coupling on other
reconstruction techniques for compressed sensing.
For example, Donoho, Maleki and Montanari recently proposed an {\em
approximate message passing} (AMP) algorithm for compressed sensing. They showed
that the AMP, when tuned to the minimax thresholding function, achieves the same
sparsity to sampling trade-off as $\ell_1$ decoding. It would be
interesting to investigate the effect of spatial coupling on AMP, especially
if it can be tuned to $\ell_p$ minimization for $0<p<1$.

\item There are also interesting open questions regarding EXIT-like curves
for compressed-sensing reconstruction.  What is the meaning, if there is any, of
the limiting curve in Figure~\ref{fig:exit_4_8}?  Can we define an EXIT-like
curve for CS reconstruction that obeys an Area Theorem like in the case of channel
coding \cite{RiU08}.

\end{enumerate}

\section{Acknowledgments} SK acknowledges support of NMC via the NSF
collaborative grant CCF-0829945 on ``Harnessing Statistical Physics for
Computing and Communications''.  The work of HP was supported by the National
Science Foundation under Grant No. CCF-0747470.  The authors would also like to thank
Misha Chertkov, Andrea Montanari, R\"udiger Urbanke 
for many useful discussions and encouragement.

\bibliographystyle{IEEEtran}
\bibliography{WCLabrv,WCLnewbib,lth,lthpub}

\end{document}